\documentclass{iopart}

\usepackage{amssymb}

\begin{document}

\title[Measure of phonon-number moments and motional quadrature]{Measure of phonon-number moments and motional quadratures through
infinitesimal-time probing of trapped ions}

\date{\today}
\author{T Bastin$^1$, J von Zanthier$^2$ and E Solano$^3$}
\address{$^1$Institut de Physique Nucl\'eaire, Atomique et de
Spectroscopie, Universit\'e de Li\`ege au Sart Tilman, B\^at.\ B15,
B - 4000 Li\`ege, Belgium}
\address{$^2$Institut f\"ur Optik,
Information und Photonik, Universit\"at Erlangen-N\"urnberg,
Staudtstrasse 7/B 2, D - 91058 Erlangen,
Germany}
\address{$^3$Max-Planck-Institut f\"ur Quantenoptik, Hans-Kopfermann-Strasse 1, D
- 85748 Garching, Germany \\
Secci\'on F\'isica, Departamento de Ciencias, Pontificia Universidad
Cat\'olica del Per\'u, Apartado 1761, Lima, Peru}
\eads{\mailto{T.Bastin@ulg.ac.be},
\mailto{jvz@optik.uni-erlangen.de},
\mailto{enrique.solano@mpq.mpg.de}}

\begin{abstract}
A method for gaining information about the phonon-number moments
and the generalized nonlinear and linear quadratures in the motion
of trapped ions (in particular, position and momentum) is
proposed, valid inside and outside the Lamb-Dicke regime. It is
based on the measurement of first time derivatives of electronic
populations, evaluated at the motion-probe interaction time $\tau
= 0$. In contrast to other state-reconstruction proposals, based
on measuring Rabi oscillations or dispersive interactions, the
present scheme can be performed resonantly at infinitesimal short
motion-probe interaction times, remaining thus insensitive to
decoherence processes.
\end{abstract}



\section{Introduction}

A single ion in a radio-frequency trap is an ideal system for the
investigation of the basic and intriguing features of quantum
mechanics~\cite{Lei03}. Recent advances in the manipulation of the
internal and external degrees of freedom of the trapped particle
allowed, for example, the realization of laser cooling to the
ground state for extended periods of time~\cite{King98,Roo99}, the
creation of nonclassical motional states~\cite{Mee96,Lei96},
entanglement between external and internal degrees of freedom of
one ion or several ions~\cite{Mon96,Tur98,Sac00,Roo04,Roo04a},
quantum gates for quantum computation~\cite{Monr95,Gul03,Sch03},
and single photon emission on demand~\cite{Kel04}.

An important aspect at this advanced level of quantum state
engineering is the ability to obtain as much information as
possible, and with high efficiency, about the motional state of
the trapped atom. Moreover, since the Lamb-Dicke (LD) regime,
where the motion of the particle is restricted to a region smaller
than a wavelength, is sometimes difficult to achieve in a trap, it
is desirable to have measurement schemes working also outside this
regime~\cite{Vog95}.

So far, several techniques have been proposed to gain information
about the motional quantum state of a trapped particle. The
central idea in all these methods is the mapping of the ionic
external dynamics to an internal degree of freedom, where the
latter can be read out, for example, with electron shelving
techniques~\cite{Deh75}. Some of these procedures are applicable
only inside the LD regime~\cite{Cir94,Wal95,Hel95,Hel96}, whereas
others work also beyond this
limit~\cite{Poy96,Bar96,Wal97,Mat98,Wal99,Sol99}, allowing to
derive in some cases the complete motional density operator.
However, a typical requirement of conventional schemes is that the
measurements have to be performed during relatively long
interaction times, through several Rabi oscillations in the
resonant case or slow phase shifts in the dispersive case, where
decoherence mechanisms cannot be neglected.

In this paper, we present a method that allows to determine the
expectation value of the phonon-number moments $\langle \hat{n}^p
\rangle$ of a trapped ion for any positive integer $p$, as well as
any generalized motional nonlinear quadrature $\frac{1}{2} \langle
\hat{f}(\hat{n}) \hat{a} e^{-i\phi} + \hat{a}^{\dagger}
\hat{f}(\hat{n}) e^{i \phi} \rangle$~\footnote{Generalized nonlinear
quadratures can be related to nonlinear coherent states,
theoretically studied in~\cite{Man96}. A method for generating
nonlinear coherent states in trapped ions, but not for measuring
them, can be found in~\cite{Mat96}.}. Here, $\hat{a}$ and
$\hat{a}^{\dagger}$ are the phonon annihilation and creation
operators, respectively, $\hat{n} = \hat{a}^{\dagger}\hat{a}$,
$\hat{f}(\hat{n})$ is a function of $\hat{n}$, and $\phi$ is an
arbitrary phase factor. In particular, we are able to measure the
expectation value of the linear quadratures, i.e., when
$\hat{f}(\hat{n}) = 1$, including the ionic position and momentum.
Our technique is not restricted to the LD regime and, moreover, can
be extended easily to all spatial dimensions, or to systems
containing more than one trapped ion. Furthermore, in contrast to
the methods presented so far, our proposal allows to obtain the
required information in an extremely short motion-probe interaction
time, which is useful in particular in presence of strong
decoherence processes.

In Sec.~\ref{singleIonCase}, we show for a single trapped ion how
to obtain information about the expectation values of the
phonon-number moments (Sec.~\ref{phonon-number moments}) and the
generalized motional nonlinear and linear quadratures
(Sec.~\ref{quadratures}) through infinitesimal-time motion-probe
interactions. In Sec.~\ref{NIonCase}, we extend the method to the
$N$-ion case. In Sec.~\ref{Summary}, we summarize our central
results.

\section{Single ion case}
\label{singleIonCase}

\subsection{Phonon-number moments}
\label{phonon-number moments}

We consider a single two-level ion trapped in a harmonic
potential. For the sake of simplicity, we confine our treatment to
one motional degree of freedom, though all results are easily
generalizable to three dimensions. The system is described by the
Hamiltonian
\begin{equation}
\hat{H}_0 = \hbar \nu(\hat{n} + 1/2) + \hbar \omega_0
|e\rangle\langle e|,
\end{equation}
where $\nu$ is the harmonic oscillator frequency, $|e\rangle$ and
$|g\rangle$ are the electronic upper and lower states of the
two-level ion, respectively, and $\omega_0$ is the associated
transition frequency.

The ion is excited by a travelling laser beam resonant with the
carrier electronic transition, leaving unchanged the motional
populations. In the usual rotating-wave approximation, and in the
interaction picture, the Hamiltonian reads~\cite{Vog95,Mat98}
\begin{equation}
\label{HintOneLaser}
\hat{H}_{\mathrm{int}} = \frac{1}{2}\hbar \Omega_L
(\hat{\sigma}^+ + \hat{\sigma}^-) \hat{f}_0(\hat{n};\eta),
\end{equation}
where $\Omega_L$ is the Rabi frequency, $\hat{\sigma}^{\pm}$ are
the electronic two-level flip operators, $\eta$ is the LD
parameter ($\eta = k_x \langle x_0 \rangle$, $\langle x_0 \rangle$
being the extension of the ground state of the motional mode and
$k_x$ the projection of the laser wave vector on the trap axis),
and $\hat{f}_0(\hat{n};\eta)$ is given by
\begin{equation}
\label{f0}
\hat{f}_0(\hat{n};\eta) = e^{-\eta^2/2}
\sum_{l=0}^{\infty} \frac{(i \eta)^{2 l}}{l!^2}
\frac{\hat{n}!}{(\hat{n} - l)!}.
\end{equation}

At any time $t$, the probability of finding the ion in the excited state
$|e\rangle$ is given by
\begin{equation}
P_e(t) = \mathrm{Tr}\Big[\hat{\rho}(t) |e\rangle\langle e|\Big] =
\Big\langle |e\rangle \langle e| \Big\rangle,
\end{equation}
where $\hat{\rho}(t)$ is the system density operator, describing
the internal and external degrees of freedom of the trapped
particle.

Considering that, for any operator $\hat{A}$,
\begin{equation}
\frac{d}{dt} \langle \hat{A} \rangle = \frac{1}{i \hbar}
\Big\langle [\hat{A},\hat{H}] \Big\rangle + \Big\langle
\frac{\partial \hat{A}}{\partial t} \Big\rangle,
\end{equation}
we get, for $\hat{A} = | e \rangle \langle e |$,
\begin{equation}
\label{dPedt} \frac{d P_e}{dt} = \frac{1}{i
\hbar} \Big\langle \Big[ |e\rangle\langle e| , \hat{H} \Big]
\Big\rangle.
\end{equation}
In our case, we have $\big[ |e\rangle\langle e| , \hat{H}_0 \big] =
0$, so that
\begin{equation} \label{dPedtA}
\frac{d P_e}{dt} = \frac{1}{i \hbar} \Big\langle \Big[
|e\rangle\langle e| , \hat{H}_{\mathrm{int}} \Big] \Big\rangle.
\end{equation}
Since, in the interaction picture,
\begin{equation}
\label{eeHint} \big[ |e\rangle\langle e| , \hat{H}_{\mathrm{int}}
\big] = \frac{\hbar \Omega_L}{2}(\hat{\sigma}^+ - \hat{\sigma}^-)
\hat{f}_0(\hat{n}, \eta),
\end{equation}
we obtain, from (\ref{dPedtA}) and (\ref{eeHint}),
\begin{equation}
\label{dPedt2} \frac{d P_e}{dt} = \frac{\Omega_L}{2 i}\mathrm{Tr}
\big[ \hat{\rho}(t) (\hat{\sigma}^+ - \hat{\sigma}^-)
\hat{f}_0(\hat{n}, \eta) \big].
\end{equation}

Next, we consider that the ion is prepared initially in the state
\begin{equation}
\hat{\rho}(0) = |\pm_\phi\rangle \langle \pm_\phi| \otimes
\hat{\rho}_f,
\end{equation}
where
\begin{equation}
\label{statepmphi} |\pm_{\phi}\rangle =
\frac{1}{\sqrt{2}}(|g\rangle \pm e^{i \phi} |e\rangle) ,
\end{equation}
and $\hat{\rho}_f$ is the phonon state we aim to characterize.
Then, a straightforward calculation yields
\begin{equation}
\label{trace1ion} \mathrm{Tr} \big[ \hat{\rho}(0) (\hat{\sigma}^+
- \hat{\sigma}^-) \hat{f}_0(\hat{n}, \eta) \big] = \mp i
\sin(\phi) \mathrm{Tr} \big[ \hat{\rho}_f \hat{f}_0(\hat{n}, \eta)
\big] ,
\end{equation}
and thus, using Eq.~(\ref{dPedt2}),
\begin{equation}
\label{dPedt3} \left. \frac{d
P^{\pm_{\phi}}_e}{d\tau}\right|_{\tau = 0} = \mp \sin(\phi)
\langle \hat{f}_0(\hat{n}, \eta) \rangle,
\end{equation}
where $\tau$ is the dimensionless time $\Omega_L t/2$ and the
token $\pm_{\phi}$ stands for the two parameters (sign $\pm$ and
phase $\phi$) defining state $| \pm_{\phi} \rangle$ of
Eq.~(\ref{statepmphi}).

Eq.~(\ref{dPedt3}) shows that the mean value of the nonlinear
operator $\hat{f}_0(\hat{n};\eta)$ is determined by the time
derivative of the population of the excited state at the initial
interaction time $\tau = 0$. This allows to gain information about
the ionic motional state, in particular, as will be shown below,
about the phonon-number moments. Moreover, since the time derivative
of the excitation probability $P^{\pm_{\phi}}_e$ is evaluated at
$\tau =0$, this information can be obtained in a very short
motion-probe interaction time, even before decoherence mechanisms
can affect the initial coherent evolution. Remark further that the
``contrast'' in the measurement of $\langle \hat{f}_0(\hat{n}, \eta)
\rangle$, as seen in Eq.~(\ref{dPedt3}), can be tuned with the
proper choice of the phase $\phi$.

For small LD parameters, a series expansion of the
nonlinear operator $\hat{f}_0(\hat{n};\eta)$ leads to the
expression
\begin{eqnarray}
\label{measure} \langle \hat{f}_0(\hat{n};\eta) \rangle \simeq 1 -
\eta^2 \langle \hat{n} \rangle + \frac{\eta^4}{4} \langle
\hat{n}^2 \rangle.
\end{eqnarray}

Thus, if we repeat the measurements with two known LD parameters,
$\eta_1$ and $\eta_2$ (varying, e.g., the angle between the laser
beam and the trap axis), we can derive $\langle\hat{n}\rangle$ and
$\langle\hat{n}^2\rangle$ by solving the linear system
\begin{equation}
\renewcommand{\arraystretch}{1.8}
\left\{
\begin{array}{rl} \langle \hat{f}_0(\hat{n}, \eta_1) \rangle & = 1 -
\eta_1^2 \langle\hat{n}\rangle + \frac{\eta_1^4}{4}
\langle\hat{n}^2\rangle, \\
\langle \hat{f}_0(\hat{n}, \eta_2) \rangle & = 1 - \eta_2^2
\langle\hat{n}\rangle + \frac{\eta_2^4}{4} \langle\hat{n}^2 \rangle,
\end{array}
\right.
\renewcommand{\arraystretch}{1}
\label{ldparameters}
\end{equation}
from which, for example, the Fano-Mandel parameter \cite{Man95}
\begin{equation}
Q = \frac{\langle \hat{n}^2 \rangle - \langle \hat{n} \rangle^2}
{\langle \hat{n} \rangle}
\end{equation}
can be extracted. It is known that the Fano-Mandel parameter allows
to determine the ``classicality'' of a given motional state, since
quantum states with $Q > 1$ are considered as classical states, and
those with $Q < 1$ as nonclassical states, due to their
sub-Poissonian phonon-number distribution.

This low order moment determination scheme can be generalized to
higher order vibronic moments $\langle \hat{n}^p \rangle$ ($p>0$).
Indeed, for higher LD parameters, the series expansion of
$\hat{f}_0(\hat{n}, \eta)$ in Eq.~(\ref{f0}) will extend beyond the
$\langle \hat{n}^2 \rangle$ term, up to a certain moment $\langle
\hat{n}^N \rangle$~\footnote{The series of Eq.~(\ref{f0}) is
convergent for all $\eta$. When the sum of this series is truncated
at $l=N$, the coefficient of the $\hat{n}^N$ term is given by
$(-1)^N e^{-\eta^2/2} {\eta}^{2N}/N!^2$. For any $\eta$ value, this
coefficient can be made arbitrarily small by a proper choice of $N$
and the series may be truncated within good approximation up to this
term.}. We could repeat then the measurement technique, as in
Eq.~(\ref{ldparameters}), for $N$ known LD parameters, yielding a
linear system from which the $N$ first vibronic moments become
accessible. Each $\langle \hat{n}^p \rangle$, $0 < p \leq N$, could
also be determined by measuring only once the initial time
derivative of $P^{\pm_{\phi}}_e$. To show this, we require that the
ion is submitted to $N$ simultaneous laser interactions, each of
them resonant with the electronic transition and leaving the
motional state unchanged. In the interaction picture, they give rise
to the simultaneous action of Hamiltonians
\begin{equation}
\hat{H}_{\mathrm{int}}^j = \frac{1}{2}\hbar \Omega_j
\hat{\sigma}^+ \hat{f}_0(\hat{n};\eta_j) + {\rm H.c.}, \quad j =
1, \ldots, N,
\end{equation}
where $\Omega_j$ and $\eta_j$ are, respectively, the electronic
Rabi frequency and the LD parameter of laser $j$. The total
Hamiltonian will be given in this case by
\begin{equation}
\label{HintNlaser} \hat{H}_{\mathrm{int}} = \sum_{j=1}^N
\hat{H}_{\mathrm{int}}^j = \frac{\hbar \Omega_L}{2}
(\hat{\sigma}^+ + \hat{\sigma}^-) \hat{F}_0(\hat{n}),
\end{equation}
where $\Omega_L = \mathrm{max}(\Omega_j)$ and
\begin{equation}
\hat{F}_0(\hat{n}) = \sum_{j = 1}^N \frac{\Omega_j}{\Omega_L}
\hat{f}_0(\hat{n};\eta_j).
\end{equation}

If the system is again initially prepared in the state
$|\pm_{\phi}\rangle\langle\pm_{\phi}| \otimes \hat{\rho}_f$, we
get, similarly to Eq.~(\ref{dPedt3}),
\begin{equation}
\label{dPedt4} \left. \frac{d
P^{\pm_{\phi}}_e}{d\tau}\right|_{\tau = 0} = \mp \sin(\phi)
\langle \hat{F}_0(\hat{n}) \rangle.
\end{equation}

It has been shown by de Matos Filho and Vogel~\cite{Mat98} that
$\hat{F}_0(\hat{n})$ may be rewritten in the form of a Taylor
series
\begin{equation}
\label{F0}
\hat{F}_0(\hat{n}) = \sum_{p=0}^{\infty} c_p \hat{n}^p,
\end{equation}
where the Taylor coefficients $c_p$ are given by
\begin{equation}
\renewcommand{\arraystretch}{1.2}
c_p = \left\{ \begin{array}{cl}
\displaystyle \sum_{j=1}^N e^{-\eta_j^2/2} \frac{\Omega_j}{\Omega_L},
& \mathrm{if} \quad p = 0, \\
\displaystyle (-1)^p \sum_{j=1}^N e^{-\eta_j^2/2}
\frac{\Omega_j}{\Omega_L} \left( \sum_{m=p}^{\infty} a_p^m
\frac{\eta_j^{2m}}{m!^2} \right), & \mathrm{if} \quad p \neq 0,
\end{array} \right.
\renewcommand{\arraystretch}{1}
\end{equation}
with
\begin{equation}
\renewcommand{\arraystretch}{1.2}
a_p^m = \left\{ \begin{array}{cl}
\displaystyle 1, & \mathrm{if} \quad p = m, \\
\displaystyle \sum_{j_{i_1} < j_{i_2} < \ldots <
j_{i_{m-p}}=1}^{m-1} j_{i_1} j_{i_2} \ldots j_{i_{m-p}} &
\mathrm{if} \quad p < m.
\end{array} \right.
\renewcommand{\arraystretch}{1}
\end{equation}

This means that for given values of the $N$ LD parameters $\eta_j$
(fixed by the geometry of the laser beams regarding to the trap
axis up to the maximum value $\eta_{\mathrm{max}} = k \langle x_0
\rangle$, $k$ being the wave vector of the lasers), the $c_p$
coefficients are linear combination of all Rabi frequencies
$\Omega_j$. In this way, the use of $N$ lasers allows to fix at
will $N$ coefficients of the Taylor series. In particular, if the
$N$ first coefficients are set to 0, we obtain
\begin{equation}
\hat{F}_0(\hat{n}) =
\mathcal{O}(\frac{\eta_{\mathrm{max}}^{2N}}{N!^2} \hat{n}^N).
\end{equation}
Note that it is always possible to choose the $N$ value such that
$\mathcal{O}(\frac{\eta_{\mathrm{max}}^{2N}}{N!^2} \hat{n}^N)$ is
negligible, even outside the LD regime where $\eta_{\mathrm{max}}$
can be greater than 1. The mean value of
$\mathcal{O}(\frac{\eta_{\mathrm{max}}^{2N}}{N!^2} \hat{n}^N)$ can
be verified experimentally by measuring the time derivative of the
population of the excited state at interaction time $\tau = 0$.
According to Eq.~(\ref{dPedt4}), the outcome of this measurement
yields $\langle \mathcal{O}(\frac{\eta_{\mathrm{max}}^{2N}}{N!^2}
\hat{n}^N) \rangle$ so that we can check if it is indeed negligible
for the phonon distribution we aim to characterize (in this case it
is recommendable to set $\phi = \pm \pi/2$).

The Rabi frequencies of the $N$ lasers could also be chosen in
such a manner that only a single coefficient $c_p$ ($p < N$) is
equal to 1, the others remaining equal to 0. In this case,
according to Eq.~(\ref{F0}) and considering that all terms
$\hat{n}^r$, $r \geqslant N$, are negligible,
\begin{equation}
\hat{F}_0(\hat{n}) = \hat{n}^p,
\end{equation}
and the measurement of $\left. \frac{d P^{\pm_{\phi}}_e}{d \tau}\right|_{\tau =
0}$ yields directly the vibronic moment $\langle \hat{n}^p
\rangle$. This step can be reproduced for any $p < N$.

It is noteworthy to mention that the knowledge of the
phonon-number moments $\langle \hat{n}^p \rangle$ for all positive
integers $p$ allows to derive the complete phonon distribution
$p(n) \equiv \langle n | \hat{\rho}_f | n \rangle$, as known from
classical statistics~\cite{Kre70}. Also, the generalization to the $3D$ case
provides measurement schemes for $\langle \hat{n}_x \rangle$,
$\langle \hat{n}_y \rangle$, $\langle \hat{n}_z \rangle$, $\langle
\hat{n}_x^2 \rangle$, $\langle \hat{n}_y^2 \rangle$, $\langle
\hat{n}_z^2 \rangle$, $\langle \hat{n}_x \hat{n}_y \rangle$,
$\langle \hat{n}_x \hat{n}_z \rangle$, $\langle \hat{n}_y
\hat{n}_z \rangle$, $\langle \hat{n}_x \hat{n}_y \hat{n}_z
\rangle$, and so on. Recently, the relevance of measuring
different photon-number moments, through quantum field homodyning,
for determining the ``classicality" of arbitrary quantum states
has been considered~\cite{Vog05}.

\subsection{Generalized nonlinear quadratures}
\label{quadratures}

Next, we show that by using a red or blue sideband excitation,
expectation values of generalized nonlinear quadrature moments,
$\frac{1}{2} \langle \hat{f}(\hat{n}) \hat{a} e^{-i\phi} +
\hat{a}^{\dagger} \hat{f}(\hat{n}) e^{i \phi} \rangle$, can be
determined using similar techniques. In particular, we have access
to the expectation values of the generalized {\it linear} quadrature
moments, among them position $\langle \hat{x} \rangle$ and momentum
$\langle \hat{p} \rangle$~\footnote{In this limit our results for
trapped ions are analogous to the ones derived for measuring field
quadratures in cavity QED setups: see~\cite{Lou05}.}. As will be
shown below, these expectation values can be obtained inside and
outside of the LD regime.

Let us consider that the ion is excited with a red-sideband
detuned laser (the blue-sideband case can be treated equally,
leading to similar results). In the usual rotating-wave
approximation, and in the interaction picture, the Hamiltonian
reads~\cite{Vog95,Mat98}
\begin{equation}
\label{HintOneLaserRed} \hat{H}_{\mathrm{int}} = \frac{i}{2}\hbar
\Omega_L \eta \hat{\sigma}^+ \hat{f}_1(\hat{n};\eta) \hat{a} +
{\rm H.c.} ,
\end{equation}
where
\begin{equation}
\hat{f}_1(\hat{n};\eta) = e^{-\eta^2/2} \sum_{l=0}^{\infty}
\frac{(i \eta)^{2 l}}{l!(l+1)!} \frac{\hat{n}!}{(\hat{n} - l)!}
\,\,\, .
\end{equation}

If the ion system is initially prepared in the state
$|\pm_{\phi}\rangle\langle\pm_{\phi}| \otimes \hat{\rho}_f$, we
obtain straightforwardly from Eq.~(\ref{dPedt})
\begin{equation}
\label{reddpedt} \left. \frac{d
P^{\pm_{\phi}}_e}{d\tau}\right|_{\tau = 0} = \pm \frac{1}{2}
\left\langle \hat{f}_1(\hat{n};\eta) \hat{a} e^{-i \phi} +
\hat{a}^{\dagger} \hat{f}_1(\hat{n};\eta) e^{i \phi} \right\rangle
,
\end{equation}
where $\tau$ is the dimensionless time $\eta \Omega_L t/2$. In the
LD regime, $\hat{f}_1(\hat{n};\eta) = 1$ and Eq.~(\ref{reddpedt})
reduces to the generalized quadrature $\hat{X}_{\phi} \equiv (
\hat{a} e^{-i \phi} + \hat{a}^{\dagger} e^{i \phi} ) /2$,
\begin{equation}
\label{reddpedtLD} \left. \frac{d
P^{\pm_{\phi}}_e}{d\tau}\right|_{\tau = 0} = \pm \langle
\hat{X}_{\phi} \rangle =  \pm \frac{1}{2} \left( \cos(\phi)
\frac{\langle \hat{x} \rangle}{\langle x_0 \rangle} + \sin(\phi)
\frac{\langle \hat{p} \rangle}{\langle p_0 \rangle} \right),
\end{equation}
where $\hat{x}$ and $\hat{p}$ are the trapped ion position and
momentum, respectively, and $\langle x_0 \rangle$ and $\langle p_0
\rangle$ are the spreads of these quantities in the ground state
of the trap potential. Eq.~(\ref{reddpedtLD}) shows that any
quadrature moment $\hat{X}_{\phi}$ can be experimentally
determined from the measurement of the initial time derivative of
the population of the excited state, whereas, in particular,
$\langle \hat{x} \rangle$ and $\langle \hat{p} \rangle$ can be
obtained by choosing $\phi = 0, \pi$ and $\phi = \pm \pi/2$,
respectively.

Outside the LD regime, a similar information is gained
using $N$ simultaneous red-sideband detuned lasers. In the
interaction picture, they give rise to the Hamiltonians
\begin{equation}
\hat{H}_{\mathrm{int}}^j = \frac{i}{2}\hbar \Omega_j \eta_j
\hat{\sigma}^+ \hat{f}_1(\hat{n};\eta_j)\hat{a} + {\rm H.c.}, \quad j =
1, \ldots, N,
\end{equation}
where $\Omega_j$ and $\eta_j$ are, respectively, the electronic
Rabi frequency and the LD parameter of laser $j$. The total
Hamiltonian is given in this case by
\begin{equation}
\hat{H}_{\mathrm{int}} = \sum_{j=1}^N \hat{H}_{\mathrm{int}}^j =
\frac{i}{2}\hbar\Omega_L \hat{\sigma}^+ \hat{F}_1(\hat{n}) \hat{a}
+ {\rm H.c.} ,
\end{equation}
where $\Omega_L = \mathrm{max}(\Omega_j)$ and
\begin{equation}
\hat{F}_1(\hat{n}) = \sum_{j=1}^N \frac{\Omega_j \eta_j}{\Omega_L}
\hat{f}_1(\hat{n} ; \eta_j).
\end{equation}

If the ion is initially in the state
$|\pm_{\phi}\rangle\langle\pm_{\phi}| \otimes \hat{\rho}_f$, we
obtain, similarly to Eq.~(\ref{reddpedt}), with $\tau = \Omega_L
t/2$,
\begin{equation}
\label{reddpedt2} \left. \frac{d
P^{\pm_{\phi}}_e}{d\tau}\right|_{\tau = 0} = \pm \frac{1}{2}
\left\langle \hat{F}_1(\hat{n}) \hat{a} e^{-i \phi} +
\hat{a}^{\dagger} \hat{F}_1(\hat{n}) e^{i \phi} \right\rangle .
\end{equation}

According to de Matos Filho and Vogel~\cite{Mat98},
$\hat{F}_1(\hat{n})$ may be again rewritten in the form of a
Taylor series
\begin{equation}
\hat{F}_1(\hat{n}) = \sum_{p=0}^{\infty} c_p \hat{n}^p,
\end{equation}
where the $c_p$ coefficients are linear combination of the $N$
laser Rabi frequencies $\Omega_j$. Similarly to the case of a
resonant laser described in the preceding section, we can choose
the $N$ value in such a way that $\mathcal{O}(c_N \hat{n}^N)$ is
negligible and fix next the $N$ Rabi frequencies to set all
coefficients $c_p$ ($p < N$) to 0, except $c_0$ to $1$. In this
case, $\hat{F}_1(\hat{n}) \simeq 1$, Eq.~(\ref{reddpedt2}) reduces
to Eq.~(\ref{reddpedtLD}) and the quadrature moments (and more
specifically $\langle \hat{x} \rangle$ and $\langle \hat{p}
\rangle$) are determined as done in the case of the LD regime.
Clearly, by following a similar procedure, we could also engineer
$\hat{F}_1(\hat{n})$ to describe an arbitrary polynomial.

\section{$N$-ion case}
\label{NIonCase}

We now briefly discuss how our proposal can be generalized to a
chain of $N$ identical two-level ions in a linear Paul trap. This
system is described by the Hamiltonian
\begin{equation}
\hat{H}_0 = \sum_{j=1}^N \hbar \nu_j(\hat{n}_j + 1/2) + \hbar
\omega_0 \sum_{k=1}^N |e\rangle_k \langle e|,
\end{equation}
where $\nu_j$ are the frequencies associated with the collective
motional modes, $\hat{n}_j = \hat{a}^{\dagger}_j\hat{a}_j$ are the
respective phonon-number operators, $|e\rangle_k$ and
$|g\rangle_k$ are the electronic states of the two-level ion $k$,
and $\omega_0$ is the electronic transition frequency of each ion.

The ions are illuminated by a laser beam resonant with their electronic
transition while leaving unchanged the phonon population,
realizing a nonlinear carrier excitation. In the usual
rotating-wave approximation and in the interaction picture, the
Hamiltonian reads~\cite{Vog95,Mat98}
\begin{equation}
\label{HintOneLaserNions} \hat{H}_{\mathrm{int}} =
\frac{1}{2}\hbar \Omega_L \sum_{k=1}^N(\hat{\sigma}^+_k +
\hat{\sigma}^-_k) \prod_{j=1}^N \hat{f}_0(\hat{n}_j;\eta_j),
\end{equation}
where $\Omega_L$ is the electronic Rabi frequency,
$\hat{\sigma}^{\pm}_k$ are the electronic two-level flip operators
of atom $k$, $\eta_j$ is the LD parameter related to the
collective mode $j$ ($\eta_j = k_x \langle x_0 \rangle_j$,
$\langle x_0 \rangle_j$ being the extension of the ground state of
mode $j$ and $k_x$ the projection of the laser wave vector on the
trap axis), and
\begin{equation}
\hat{f}_0(\hat{n}_j;\eta_j) = e^{-\eta_j^2/2} \sum_{l=0}^{\infty}
\frac{(i \eta_j)^{2 l}}{l!^2} \frac{\hat{n}_j!}{(\hat{n}_j - l)!}.
\end{equation}
In the following, we denote the product $\prod_{j=1}^N
\hat{f}_0(\hat{n}_j;\eta_j)$ by $\hat{{\cal F}}_0$. The
generalized relation
\begin{equation}
\label{dPedtk} \displaystyle \frac{d (P_e)_k}{dt} = \frac{1}{i
\hbar} \Big\langle \Big[ |e\rangle_k \langle e| , \hat{H} \Big]
\Big\rangle,
\end{equation}
can be easily obtained, as in Eq.~(\ref{dPedt}), where $(P_e)_k$
is the probability of finding ion $k$ in its excited state. As
$|e\rangle_k \langle e|$ commutes with $\hat{H}_0$ and any
operator associated with other ions $k'$, we get immediately
\begin{equation}
\label{dPedtk2} \frac{d (P_e)_k}{dt} = \frac{1}{i \hbar}
\Big\langle \Big[ |e\rangle_k \langle e| , \frac{1}{2}\hbar
\Omega_L (\hat{\sigma}^+_k + \hat{\sigma}^-_k) \hat{{\cal F}}_0
\Big] \Big\rangle,
\end{equation}
and thus
\begin{equation}
\label{dPedt2N} \frac{d (P_e)_k}{dt} = \frac{\Omega_L}{2
i}\mathrm{Tr}\big[ \hat{\rho}(t) (\hat{\sigma}^+_k -
\hat{\sigma}^-_k) \hat{{\cal F}}_0 \big],
\end{equation}
where $\hat{\rho}(t)$ is the $N$-ion system density operator.

Let us consider the following initial state
\begin{equation}
\hat{\rho}(0) = \hat{\rho}_k \otimes \hat{\rho}_A \otimes
\hat{\rho}_f,
\end{equation}
where
\begin{equation}
\hat{\rho}_k = |\pm_\phi\rangle_k \langle \pm_\phi|
\end{equation}
is the electronic density operator of the $k$-th ion, with
\begin{equation} |\pm_{\phi}\rangle_k =
\frac{1}{\sqrt{2}}(|g\rangle_k \pm e^{i \phi} |e\rangle_k),
\end{equation}
$\hat{\rho}_A$ is an arbitrary electronic density operator of the
remaining ions, and $\hat{\rho}_f$ is the collective motional
density operator associated with the $N$ eigenmodes. Following
similar steps as in the previous sections, we obtain
\begin{eqnarray}
\mathrm{Tr}\big[ \hat{\rho}(0) (\hat{\sigma}^+_k -
\hat{\sigma}^-_k) \hat{{\cal F}}_0 \big] = \mp i \sin(\phi)
\mathrm{Tr} \big[ \hat{\rho}_f \hat{{\cal F}}_0 \big],
\end{eqnarray}
similar to Eq.~(\ref{trace1ion}), and thus, using
Eq.~(\ref{dPedt2N}),
\begin{equation}
\label{dPedtk3} \left. \frac{d (P_e)_k}{d\tau}\right|_{\tau = 0} =
\mp \sin(\phi) \langle \hat{{\cal F}}_0 \rangle,
\end{equation}
where $\tau$ is the dimensionless time $\tau = \Omega_L t/2$.
Eq.~(\ref{dPedtk3}) tells us that by measuring the time derivative
of the population of the excited state of {\it one} ion at $\tau =
0$, we are able to gain information about the {\it collective}
nonlinear operator $\hat{{\cal F}}_0$.

\section{Summary}

\label{Summary}

In conclusion, we have proposed a method that allows to determine
the phonon-number moments and the motional nonlinear and linear
quadratures of a trapped ion, in particular the ionic position and
momentum. Our method makes use of the nonlinear behavior of the
ion-laser interaction in harmonic traps. The measurement of the
phonon-number moments $\langle \hat{n}^p \rangle$ requires
resonant carrier interaction, with no phonon gain or loss, while
the measurement of the nonlinear quadrature moments $\frac{1}{2}
\langle \hat{f}(\hat{n}) \hat{a} e^{-i\phi} + \hat{a}^{\dagger}
\hat{f}(\hat{n}) e^{i \phi} \rangle$ demands the use of red or
blue sideband excitations. In contrast to methods presented so
far, our proposal is designed for measurements realized in very
short probe laser interaction times, thus preventing the noisy
action of decoherence processes. In addition, we have shown that
our scheme works inside and outside the LD regime, and that it can
be generalized to the case of $N$ ions. In the latter case,
information about a collective property is gained by the
measurement of the excited state population of a single ion.

\ack
TB acknowledges support from the Belgian Institut
Interuniversitaire des Sciences Nucl\'eaires (IISN), and thanks
JVZ and ES for the hospitality at the University of Erlangen,
Erlangen, and Max-Planck-Institut f\"ur Quantenoptik, Garching,
Germany. ES acknowledges support from EU through RESQ project.

\end{document}